%% file: CMDM.tex
\documentclass[12pt,epsf]{article}
\usepackage{amsmath,amssymb,graphicx}
\usepackage{epsfig}
\usepackage{comment}

\begin{document}

\def\a{\alpha}
\def\b{\beta}
\def\c{\varepsilon}
\def\d{\delta}
\def\e{\epsilon}
\def\f{\phi}
\def\g{\gamma}
\def\h{\theta}
\def\k{\kappa}
\def\l{\lambda}
\def\m{\mu}
\def\n{\nu}
\def\p{\psi}
\def\q{\partial}
\def\r{\rho}
\def\s{\sigma}
\def\t{\tau}
\def\u{\upsilon}
\def\v{\varphi}
\def\w{\omega}
\def\x{\xi}
\def\y{\eta}
\def\z{\zeta}
\def\D{\Delta}
\def\G{\Gamma}
\def\H{\Theta}
\def\L{\Lambda}
\def\F{\Phi}
\def\P{\Psi}
\def\S{\Sigma}

\def\o{\over}
\def\beq{\begin{eqnarray}}
\def\eeq{\end{eqnarray}}
\newcommand{\gsim}{ \mathop{}_{\textstyle \sim}^{\textstyle >} }
\newcommand{\lsim}{ \mathop{}_{\textstyle \sim}^{\textstyle <} }
\newcommand{\vev}[1]{ \left\langle {#1} \right\rangle }
\newcommand{\bra}[1]{ \langle {#1} | }
\newcommand{\ket}[1]{ | {#1} \rangle }
\newcommand{\EV}{ {\rm eV} }
\newcommand{\KEV}{ {\rm keV} }
\newcommand{\MEV}{ {\rm MeV} }
\newcommand{\GEV}{ {\rm GeV} }
\newcommand{\TEV}{ {\rm TeV} }
\def\diag{\mathop{\rm diag}\nolimits}
\def\Spin{\mathop{\rm Spin}}
\def\SO{\mathop{\rm SO}}
\def\O{\mathop{\rm O}}
\def\SU{\mathop{\rm SU}}
\def\U{\mathop{\rm U}}
\def\Sp{\mathop{\rm Sp}}
\def\SL{\mathop{\rm SL}}
\def\tr{\mathop{\rm tr}}

\def\IJMP{Int.~J.~Mod.~Phys. }
\def\MPL{Mod.~Phys.~Lett. }
\def\NP{Nucl.~Phys. }
\def\PL{Phys.~Lett. }
\def\PR{Phys.~Rev. }
\def\PRL{Phys.~Rev.~Lett. }
\def\PTP{Prog.~Theor.~Phys. }
\def\ZP{Z.~Phys. }


\baselineskip 0.7cm

\begin{titlepage}

\begin{flushright}
UT-09-28\\
IPMU-09-0156
\end{flushright}

\vskip 1.35cm
\begin{center}
{\large \bf
   Low-Scale Gauge Mediation \\and\\ Composite Messenger Dark Matter
}
\vskip 1.2cm
Koichi Hamaguchi$^{1,2}$, Eita Nakamura$^{1,2}$, Satoshi Shirai$^{2,1}$, and Tsutomu T. Yanagida$^{2,1}$
\vskip 0.4cm

{\it $^1$ Department of Physics, University of Tokyo,\\
     Tokyo 113-0033, Japan \\
$^2$ Institute for the Physics and Mathematics of the Universe, 
University of Tokyo,\\ Chiba 277-8568, Japan}

\vskip 1.5cm

\abstract{
In recent works, we have proposed a possible dark matter in composite messenger gauge mediation models.
In this paper, we discuss the details of a composite messenger model taking a possible supersymmetry breaking
scenario and show that the correct dark matter abundance and a successful gauge
mediation can be realized. 
}
\end{center}
\end{titlepage}

\setcounter{page}{2}

\section{Introduction}

The origin of dark matter (DM) in the Universe is still not identified,
although there have been accumulated numerous observations for the existence of the DM itself. 
Therefore, it is the most important problem in particle physics and astrophysics
to determine the nature of the DM.

The DM must be absolutely stable or its lifetime should be much longer than the age of the Universe.
This least requirement may be satisfied if the DM is ultra light or if the DM has
a charge of some unbroken additional symmetry. The former example is the invisible axion of 
mass $m_a = {\cal O}(10^{-5})$ eV. The lifetime of the axion is proportional to $m_a^{-5}$ and its 
longevity is explained by the extremely small mass.
A familiar example for the latter case is the lightest 
supersymmetric (SUSY) particle (LSP) in the SUSY standard model. In this model one may impose
an R-parity conservation to guarantee the stability of the LSP.

Besides the above two categories, we have a well-established mechanism to guarantee a longevity
of a particle, that is, the compositeness. The proton is known as a QCD bound state of 
three quarks $q$. The lowest dimensional operators to destabilize the proton are dimension six
operators such as $qqql$. Those dimension six operators are suppressed by $1/M_{\rm PL}^2$, which
leads naturally to the long lifetime of the proton. Here, $M_{\rm PL}\simeq 2.4\times 10^{18}$ GeV
is the reduced Planck scale. It is very remarkable that the lifetime of the proton would be longer than
the age of the Universe even if the mass of the proton was at ${\cal O}(100)$ TeV. Thus, composite baryons 
$X$ of some new strong interactions below the ${\cal O}(100)$ TeV can be interesting candidates for the DM.

Furthermore the composite baryon and anti-baryon may annihilate into mesons with the
strong interactions and the annihilation cross section may be close to the unitarity bound.
If it is the case, one may naturally explain the present energy density of the DM for the mass
of the composite baryon being ${\cal O}(100)$ TeV \cite{Griest:1989wd,Dimopoulos:1996gy}.

Surprisingly, the above mass scale coincides with the mass scale of messengers in low-scale gauge mediations.
We have recently proposed a composite messenger model, motivated by the above coincidence of 
two independent mass scales, the masses for DM baryons and for the messengers \cite{Hamaguchi:2007rb,Hamaguchi:2008rv}.
The low-scale gauge mediations also accommodate a light gravitino of mass $m_{3/2}<{\cal O}(10)$ eV
and such a light gravitino is free from all cosmological problems \cite{G1,G2}.

The purpose of this paper is to discuss details of the composite messenger model taking a
possible SUSY breaking scenario. We find an allowed parameter range for the masses of the composite
baryon and anti-baryon together with those of the messenger composite mesons.
A detailed analysis of the model and its phenomenological consequences will be presented in a forthcoming paper \cite{NS}.

\section{A composite messenger model and a composite baryon DM}

\subsection{A composite messenger model}

Before specifying a SUSY-breaking sector, we first discuss the composite messenger model which is based on
a confining gauge theory with a gauge group $G$. We introduce messenger quarks $Q^i$ and anti-quarks ${\bar Q}_i$, both of
which transform as non-trivial representations of the gauge group $G$. Here, $i$ denotes indices of the
standard model (SM) gauge groups. In this paper we consider $G={\rm SU}(N)$ and  $Q^i$ and ${\bar Q}_i$
are ${\bf 5}$ and ${\bf 5}^*$ of the GUT ${\rm SU}(5)_{\rm GUT}$, respectively. Thus, $i$ runs from 1 to 5.
The quark $Q^i$ and anti-quark ${\bar Q}_i$ are assumed to transform as fundamental and anti-fundamental
representations of the ${\rm SU}(N)$, respectively. We omit the indices of the ${\rm SU}(N)$ throughout this paper.

We have composite baryons $Q^iQ^jQ^k\cdots$ and anti-baryons ${\bar Q}_i{\bar Q}_j{\bar Q}_k\cdots$ as well as
mesons such as $M^i_j=Q^i{\bar Q}_j$. 
Those baryons are nothing but the DM candidate. However, they carry, in general,
non-vanishing charges of the unbroken SM gauge groups, which are not suitable for the DM in the Universe.
The requirement for baryons to be neutral under the unbroken SM gauge groups leads us to consider the case of
$N$ being a multiple of 5. Here, we take the minimal case, that is ${\rm SU}(5)$, for the gauge group $G$. Then,
we have a pair of a baryon $B=Q^1Q^2Q^3Q^4Q^5$ and an anti-baryon 
${\bar B}={\bar Q}_1{\bar Q}_2{\bar Q}_3{\bar Q}_4{\bar Q}_5$.

We now introduce a mass for the messenger quarks as
\begin{equation}
W= mQ^i{\bar Q}_i.
\end{equation}
The low-energy dynamics is described by the above baryons $B,{\bar B}$ and the mesons $M^i_j$ with
the effective superpotential \cite{Seiberg:1994bz}
\begin{equation}
W_{\rm eff}= X\left({\rm det}\,M - {\bar B}B - \Lambda^{10}\right) + m\,{\rm tr}\,M,
\end{equation}
where $X$ is a Lagrange's multiplier and $\Lambda$ is the dynamically generated holomorphic scale of the gauge group $G$.
These hadrons are massless fields which describe the effective theory on the moduli space of vacua
in the limit $m\to0$.
We rescale the baryon and meson chiral superfields so that they have canonically normalized K\"ahler
potentials. After the rescaling we obtain
\begin{equation}\label{eff_superpotential}
W_{\rm eff}= \frac{g_X}{16\pi^2}X\left(\frac{g_M^5}{\Lambda^3}{\rm det}\,M - g_B^2{\bar B}B - 
\Lambda^{2}\right) + 
\frac{g_M}{16\pi^2}\Lambda m\,{\rm tr}\,M,
\end{equation}
where the couplings $g_X,g_M$ and $g_B$ are constants of ${\cal O}(4\pi)$. Here, we have used the naive dimensional
analysis (NDA) \cite{Luty}. In order to assure the analysis to be valid, we assume that $m\lesssim\Lambda$ here and after.

The above effective superpotential yields a supersymmetric vacuum as
\begin{equation}\label{SUSY_VEV}
\langle g_MM^i_j\rangle = \Lambda\delta^i_j,~~\langle B\rangle=\langle{\bar B}\rangle =0,~~\langle g_XX\rangle=-m.
\end{equation}
Notice that the theory possesses a global ${\rm SU}(5)_{\rm L}\times {\rm SU}(5)_{\rm R}$ symmetry, in the limit of $m\to 0$,
under which the meson $M^i_j$ transforms as
\begin{equation}
M\to M'=U_{\rm L}^\dagger MU_{\rm R},\quad\text{for $U_{\rm L,R}\in{\rm SU}(5)$}.
\end{equation}
Then, the meson condensation $\langle g_M M^i_j \rangle = \Lambda \delta^i_j$
spontaneously breaks the global symmetry down to the diagonal ${\rm SU}(5)_V$, 
and hence the adjoint meson excitations can also be represented as 
the (pseudo) Nambu--Goldstone (NG) superfields associated to the broken symmetry,
which we call ${\rm SU}(5)_A$.
We will use this fact
when we calculate the annihilation cross section of the DM in subsection \ref{DM_abundance}.

By substituting these vacuum expectation values (VEVs) in Eq.~(\ref{eff_superpotential}),
we find SUSY-invariant masses for the baryon and mesons. The baryon mass is given by
\begin{equation}
m_B=\frac{g_B^2}{16\pi^2}\cdot m.
\end{equation}
The mesons split into the ${\rm SU}(5)_{\rm GUT}$-adjoint mesons and a singlet meson.
The mass for the adjoint mesons is given by
\begin{equation}
m_{\rm ad}=\frac{g_M^2}{16\pi^2}\cdot m.
\end{equation}
The singlet meson mixes with the $X$. Their supersymmetric mass matrix is given by
\begin{equation}
\frac{1}{16\pi^2}\begin{bmatrix}
-g_M^24m & g_Xg_M\sqrt{5}\Lambda\\
g_Xg_M\sqrt{5}\Lambda & 0 \end{bmatrix}.
\end{equation}

Strictly, the ${\rm SU}(5)_{\rm GUT}$ group is broken down to the SM gauge group
${\rm SU}(3)\times{\rm SU}(2)\times{\rm U}(1)$ and the messengers split into the ``down-type'' and
``lepton-type'' ones. These two types of messengers have different masses $m_d$ and $m_\ell$, respectively.
For example, if we assume a common mass for these messengers at the GUT scale,
the mass ratio $m_d/m_\ell$ at the messenger mass scale
is $2.5$--$3.5$, as we see by solving the renormalization group equations.
However, this mass splitting does not alter the essential part of our discussion.
In this paper, for simplicity, we only consider the ${\rm SU}(5)_{\rm GUT}$-invariant mass,
$m_d=m_\ell=m$, at the messenger mass scale.
We will consider effects of the mass splitting in the following paper \cite{NS}.

\subsection{SUSY-breaking mass spectrum in the messenger sector}

For an explicit calculation we adopt a dynamical SUSY-breaking model in ref. \cite{Fujii:2003iw}.
The SUSY-breaking effect at low energies can be represented by a singlet field $S$
with a superpotential interaction
\begin{equation}
W\supset \frac{f}{3}S^3
\end{equation}
and a tachyonic mass term of $S$ in the scalar potential
\begin{equation}\label{neg_mass}
V\supset-\mu^2|S|^2,
\end{equation}
which breaks the supersymmetry explicitly.
In ref. \cite{Fujii:2003iw}, an explicit model which induces $\mu$ dynamically is presented.
Here, we assume that $\mu$ is induced by some unknown SUSY-breaking sector.
Due to the negative mass squared, the scalar component of $S$ acquires a nonzero VEV,
\begin{equation}\label{scalar_S}
{}|\langle S\rangle|=\frac{\mu}{\sqrt{2}|f|},
\end{equation}
which in turn results in a nonzero $F$-component of $S$:
\begin{equation}\label{F_S}
{}|\langle F_S\rangle|=\frac{\mu^2}{2|f|}.
\end{equation}

Let us introduce a coupling of the $S$ to the messenger quarks, that is,
\begin{equation}\label{yukawa_coupling}
W_{\rm coupl} = hSQ^i{\bar Q}_i.
\end{equation}
Notice that although we do not include an explicit mass term in the superpotential, the nonzero vacuum expectation
value of the $S$ induces a mass $m=h\langle S\rangle$.

First, we discuss the SUSY-breaking effect in the messenger sector using the quark picture.
By examining the scalar potential, we see that if $|f|<|h|$ there is a global minimum
where $\langle Q\rangle=\langle\bar{Q}\rangle=0$ and the scalar- and $F$-component VEVs of $S$
are given by Eqs.~(\ref{scalar_S}) and (\ref{F_S}). These nonzero VEVs induce the messengers
both holomorphic and SUSY-breaking soft masses. The condition $|f|<|h|$ is equivalent
to the stability condition that the messenger scalars have non-tachyonic masses.

Now we discuss the SUSY-breaking effect in the effective theory.
The effective theory is described by the baryon, mesons, and light singlets including the $S$.
In this hadron picture, the ${\rm SU}(5)_{\rm GUT}$-adjoint meson takes a role as the messenger field.
The effective superpotential is given by
\begin{equation}\label{eff_supot_break}
W_{\rm eff}=\frac{g_X}{16\pi^2}X\left(\frac{g_M^5}{\Lambda^3}{\rm det}\,M-g_B^2{\bar B}B-\Lambda^{2}\right)
+\frac{g_M}{16\pi^2}h\Lambda S\,{\rm tr}\,M+\frac{f}{3}S^3,
\end{equation}
and the effect of the SUSY-breaking sector is incorporated as the negative soft mass-squared for $S$ in Eq.~(\ref{neg_mass}). 
The scalar potential is given by
\begin{align}\label{scalar_pot}
V&=\left|fS^2+\frac{g_M}{16\pi^2}h\Lambda\,{\rm tr}\,M\right|^2+\left|
\frac{g_X}{16\pi^2}\left(\frac{g_M^5}{\Lambda^3}{\rm det}\,M-g_B^2B\bar{B}-\Lambda^2\right)\right|^2 \notag\\
&\quad+\left|\frac{g_X}{16\pi^2}g_B^2X\bar{B}\right|^2+\left|\frac{g_X}{16\pi^2}g_B^2XB\right|^2 \notag\\
&\quad+\sum_{i,j}\left|\frac{g_X}{16\pi^2}X\frac{g_M^5}{\Lambda^3}[{\rm Cof}\,M]_i^j+\frac{g_M}{16\pi^2}
hS\Lambda \delta_i^j\right|^2-\mu^2|S|^2.
\end{align}

We first assume that $\langle X\rangle\ne0$. Then $\langle B\rangle=\langle \bar{B}\rangle=0$ provided that
a non-tachyonic condition is satisfied. Then $\langle X\rangle$ is determined as
\begin{equation}\label{scalar_rel}
\langle X\rangle=-\frac{1}{g_X}\frac{h}{v^4}\langle S\rangle,
\end{equation}
where we assumed
\begin{equation}
\langle M_j^i\rangle=v\frac{\Lambda}{g_M}\delta_j^i.
\end{equation}
Next, the reduced scalar potential in terms of $v$ and $S$ is
\begin{equation}
V(v,S)=\left|fS^2+5\frac{h\Lambda^2}{16\pi^2}v\right|^2
+\left|\frac{g_X}{16\pi^2}\Lambda^2(v^5-1)\right|^2-\mu^2|S|^2
\end{equation}
and it can be minimized with respect to $S$ as
\begin{equation}
{}|\langle S\rangle|^2=\left|\frac{5}{f}\frac{h\Lambda^2}{16\pi^2}v\right|
+\frac{\mu^2}{2|f|^2}.
\end{equation}
Then the resulting reduced potential for $v$ is
\begin{equation}\label{red_pot}
V(v)=\left|\frac{g_X}{16\pi^2}\Lambda^2(v^5-1)\right|^2-\left|\frac{5\mu^2}{f}\frac{h\Lambda^2}{16\pi^2}v\right|
-\frac{\mu^4}{4|f|^2}.
\end{equation}
At this vacuum, we have
\begin{align}
{}|F_S|&=\frac{\mu^2}{2|f|}, \\
{}|F_X|&=\left|\frac{g_X}{16\pi^2}\Lambda^2(v^5-1)\right|
\end{align}
and a relation
\begin{equation}\label{F_rel}
hF_S+g_Xv^4F_X=0
\end{equation}
is satisfied.
Eq.~(\ref{red_pot}) shows that the reduced potential has five vacua \cite{Witten:1982df}
 which are related by multiplications by ${\rm e}^{2\pi i/5}$'s and one of the vacua has a real positive value of $v$.
We further see that this positive $v$ satisfies $v>1$ and numerical calculations show that $v={\cal O}(1)$.
We will consider a theory around this vacuum in the following.
We have a relation
\begin{equation}
{}|F_S|^2+|F_X|^2<3m_{3/2}^2M_{\rm PL}^2,
\end{equation}
where $m_{3/2}$ is the gravitino mass.

The scalar potential Eq.~(\ref{scalar_pot}) also has a local minimum with $\langle X\rangle=0$.
It can be shown that if
\begin{equation}\label{eq:UFB}
5\left|\frac{g_M^2}{16\pi^2}h\Lambda\right|^2-\mu^2>0
\end{equation}
is not satisfied, the potential has a runaway solution with $|S|\to\infty$.
As long as the condition (\ref{eq:UFB}) is satisfied, the minimization yields $\langle S\rangle=0$ for $\langle X \rangle = 0$.
We have checked that for the parameter region we are interested in, the vacuum with $\langle X\rangle\ne0$
is the global minimum.
Thus, we will consider only this vacuum with $\langle X\rangle\ne0$ in the following discussions.

Expanding around the above VEVs, we obtain
\begin{equation}
W_{\rm eff}=W_{\rm eff}^{(0)}+W_{\rm eff}^{(1)}+W_{\rm eff}^{(2)}+W_{\rm eff}^{(3)}+\cdots,
\end{equation}
where
\begin{align}
W_{\rm eff}^{(0)}&=\frac{h\Lambda^2}{16\pi^2}\langle S\rangle\left(4v+\frac{1}{v^4}\right)
+\frac{f}{3}\langle S^3\rangle,\\
W_{\rm eff}^{(1)}&=\frac{g_X}{16\pi^2}\Lambda^2(v^5-1)\delta X
-\frac{\mu^2}{2|f|}	
\delta S, \\
W_{\rm eff}^{(2)}&=-\frac{1}{2}\frac{g_M^2}{16\pi^2}\frac{h\langle S\rangle}{v}
\Big(({\rm tr}[\delta M])^2-{\rm tr}[\delta M\delta M]\Big)+f\langle S\rangle \delta S^2 \notag\\
&\quad+\frac{g_B^2}{16\pi^2}\frac{h\langle S\rangle}{v^4}\bar{B}B
+\frac{g_Xg_M}{16\pi^2}\Lambda v^4\delta X\,{\rm tr}[\delta M]
+\frac{g_M}{16\pi^2}h\Lambda \delta S\,{\rm tr}[\delta M], \\
W_{\rm eff}^{(3)}&=\frac{1}{2}\frac{g_Xg_M^2}{16\pi^2}v^3\delta X
\Big(({\rm tr}[\delta M])^2-{\rm tr}[\delta M\delta M]\Big)-\frac{g_Xg_B^2}{16\pi^2}\delta X\bar{B}B \notag\\
&\quad-\frac{1}{6}\frac{g_M^3}{16\pi^2}\frac{h\langle S\rangle}{v^2\Lambda}
\Big(({\rm tr}[\delta M])^3+2{\rm tr}[(\delta M)^3]-3{\rm tr}[\delta M]{\rm tr}[(\delta M)^2]\Big) \notag\\
&\quad+\frac{f}{3}\delta S^3.
\end{align}

From the effective superpotential we find the mass spectrum for the messenger mesons and baryons.
Now the supersymmetry is broken, and the nonzero $F$-term of the $\delta X$
produces non-holomorphic masses, which result in splittings in the scalar masses.

The fermionic baryon mass is given by
\begin{equation}
m_B^f=\frac{g_B^2}{16\pi^2}\left|\frac{h}{v^4}\langle S\rangle\right|.
\end{equation}
The adjoint meson has a fermionic mass as
\begin{equation}
m_{\rm ad}^f=\frac{g_M^2}{16\pi^2}\left|\frac{h}{v}\langle S\rangle\right|.
\end{equation}
The singlet meson, the $\delta X$, and the $S$ mix with each other. Their fermionic masses are given by the
diagonalization of ${\bf m}_{\rm singlet}^\dagger{\bf m}_{\rm singlet}$, where
\begin{equation}
{\bf m}_{\rm singlet}=\frac{1}{16\pi^2}\begin{bmatrix}-g_M^24hv^{-1}\langle S\rangle&
g_Xg_M\sqrt{5}\Lambda v^4&g_M\sqrt{5}h\Lambda\\
g_Xg_M\sqrt{5}\Lambda v^4&0&0\\
g_M\sqrt{5}h\Lambda&0&2f\langle S\rangle\end{bmatrix}.
\end{equation}

The scalar baryon masses are given by
\begin{equation}\label{DM_mass}
(m_B^s)^2=(m_B^f)^2\pm\left|\frac{g_Xg_B^2}{16\pi^2}F_X\right|,
\end{equation}
where the contribution of the second term on the right-hand side is from the nonzero $F$-term of $\delta X$.
The lighter scalar baryon is the DM candidate, which we will discuss in the next subsection.
The scalar masses of the adjoint meson are given by
\begin{equation}
(m_{\rm ad}^s)^2=(m_{\rm ad}^f)^2\pm\left|\frac{g_Xg_M^2}{16\pi^2}v^3F_X\right|.
\end{equation}
Similarly, the scalar singlet mass term is written as
\begin{equation}
V\supset \frac{1}{2}\begin{bmatrix}\varphi_{\rm singlet}^\dagger&\varphi_{\rm singlet}^{\rm T}\end{bmatrix}
R_{\rm singlet}\begin{bmatrix}\varphi_{\rm singlet}\\\varphi_{\rm singlet}^*\end{bmatrix}
\end{equation}
and the scalar singlet mass matrix is given by
\begin{equation}
R_{\rm singlet}=\begin{bmatrix}{\bf m}_{\rm singlet}^\dagger{\bf m}_{\rm singlet}&N_{\rm singlet}^\dagger\\
N_{\rm singlet}&{\bf m}_{\rm singlet}^\dagger {\bf m}_{\rm singlet}\end{bmatrix},
\end{equation}
where
\begin{equation}
N_{\rm singlet}=\begin{bmatrix}-\frac{g_M^2}{16\pi^2}4v^3 F_X&0&0\\0&0&0\\0&0&-2fF_S\end{bmatrix}.
\end{equation}

In the hadron picture, the non-tachyonic conditions for the messenger hadron masses are given by
\begin{equation}
\frac{g_M^2}{16\pi^2}\frac{|\langle S\rangle|^2}{|F_S|}>\left|\frac{v}{h}\right|
\end{equation}
for the adjoint meson mass and
\begin{equation}
\frac{g_B^2}{16\pi^2}\frac{|\langle S\rangle|^2}{|F_S|}>\left|\frac{v^4}{h}\right|
\end{equation}
for the baryon mass.

\subsection{The lightest messenger baryon as the DM and its relic abundance}\label{DM_abundance}

As described in the previous sections, the messenger baryon in our model is completely
neutral under the SM gauge group.
It has a long lifetime since the messenger baryon is a bound
state of five quarks $Q^i$'s and we have an effective baryon number
conservation as  in QCD .
Therefore the messenger baryon is a good candidate for the DM
in the Universe, provided that it has a correct abundance in the present Universe.
We show, in this subsection, that with the strong interaction of the messenger gauge group the
messenger baryon indeed has the correct abundance
in a certain region of the parameter space.

Let us first derive the constraint on the mass of the dark matter baryon \cite{Griest:1989wd}. 
In general, the unitarity limit on the S-wave annihilation cross section for the DM is given by,
\begin{eqnarray}\label{eq:unitarity}
 \sigma v_{\rm rel} \le \frac{4\pi}{m_{\rm DM}^{2} v_{\rm rel}},
\end{eqnarray}
where $m_{\rm DM}$ denotes the mass of the dark matter,  and $v_{\rm rel}$ the relative velocity
of the dark matter at the freeze out time (typically given by, $v_{\rm rel} \simeq 0.35$).
Here, we have assumed that the annihilation process is dominated by the S-wave contribution
and the DM is a complex boson since our DM is a boson component of the messenger baryon superfield (see Eq.~(\ref{DM_mass})).
The dark matter density observed by the WMAP experiment,
$\Omega_{\rm DM} h^{2}=0.1131\pm0.0034$ \cite{Komatsu:2008hk}, 
requires the annihilation cross section of the dark matter to be
\begin{eqnarray}\label{eq:csDM}
 \sigma v_{\rm rel}={\cal O}(10^{-9}) \,{\rm GeV}^{-2}.
\end{eqnarray}
Therefore, in our DM model, the current upper bound on the mass of the DM boson is given by,
\begin{eqnarray}\label{eq:UpDM}
 m_{\rm DM}\lesssim 150\,{\rm TeV}.
\end{eqnarray}

The annihilation cross sections of the DM can be estimated by a `chiral perturbation theory'-like method.
As we mentioned below Eq.~(\ref{SUSY_VEV}), the ${\rm SU}(5)_{\rm GUT}$-adjoint meson can be viewed as
a (pseudo-)NG field.
Although it seems unobvious in the effective superpotential in Eq.~(\ref{eff_supot_break}),
its interactions in the low energies are suppressed by derivatives.
To express this suppression explicitly, we use the NG-field representation
for the ${\rm SU}(5)_{\rm GUT}$-adjoint meson in the following calculation.
According to Eq.~(\ref{SUSY_VEV}), the NG field is introduced as
\begin{equation}
M={\rm exp}(\Pi/f_\pi)D\,{\rm exp}(\Pi/f_\pi),
\end{equation}
where $D$ represents the field in the directions of the ${\rm SU}(5)_{\rm GUT}$-singlets and the $\Pi$ belongs to
the adjoint representation of the ${\rm SU}(5)_{\rm GUT}$.
From the normalization of the quadratic K\"ahler potential, the decay constants are determined to be
\begin{equation}
f_\pi=4\left|v\frac{\Lambda}{g_M}\right|^2.
\end{equation}
It can be easily checked that the coupling between the meson and the $S$ field, which breaks ${\rm SU}(5)_A$ symmetry explicitly,
induces the same holomorphic and non-holomorphic masses calculated in the previous section.

Now, let us calculate the abundance of the DM baryon in terms of the NG fields.
In the above NG-field representation, the NG field disappears from the effective superpotential except
for the (explicitly) symmetry-breaking terms. Thus effective couplings between the DM and
the NG field must involve $D$-term interactions or symmetry-breaking terms. More concretely,
the effective Lagrangian does not contain terms such as
\begin{equation}
\varphi_B\varphi_{\bar{B}}({\rm tr}[\varphi_\pi\varphi_\pi])^*,
\end{equation}
and the leading contribution comes from terms involving two derivatives such as
\begin{equation}
\varphi_B\varphi_{\bar{B}}(\partial_\mu\varphi_\pi^\dagger)(\partial^\mu\varphi_\pi).
\end{equation}
Here, $\varphi_B,\varphi_{\bar{B}}$, and $\varphi_\pi$ are the scalar components of the superfields $B,\bar{B}$, and $\Pi$, 
By the NDA, coefficients of these terms are estimated as
\begin{equation}
{\cal L}\supset\frac{g_B^2g_M^2}{16\pi^2}\frac{1}{\Lambda^2}\varphi_B\varphi_{\bar{B}}(\partial_\mu\varphi_\pi^\dagger)(\partial^\mu\varphi_\pi).
\end{equation}
The annihilation cross section for the DMs, $\varphi_B$ and $\varphi_{\bar{B}}$, is given by,
\begin{equation}\label{ann_cross_sec_NG}
\sigma v_{\rm rel} \simeq \frac{d}{32\pi}\bigg(\frac{g_B^2g_M^2}{16\pi^2}\bigg)^2\frac{m_{\rm DM}^2}{\Lambda^4}
\sqrt{1-\frac{m_{\rm meson}^2}{m_{\rm DM}^2}},
\end{equation}
where $d$ is the number of the degree of freedom into which the DMs annihilate. (E.g. one real scalar particle contributes $d=1/2$.)

We also have a contribution from the explicit-breaking term. The invariant matrix element can be calculated as
\begin{equation}\label{mat_ele}
{\cal M}(B+\bar{B}\to{\rm mesons})
=\sum_{i\,:\,{\rm singlet}}\frac{g_Xg_M^2}{16\pi^2}\left[\frac{{\bf m}^\dagger_{\rm singlet}{\bf m}_{\rm singlet}}{4m_{\rm DM}^2{\bf1}
-{\bf m}^\dagger_{\rm singlet}{\bf m}_{\rm singlet}}\right]_{Xi}y_i,
\end{equation}
where $y_i$ is the Yukawa coupling between the singlet $i$ (which label the four singlets) and the mesons in the final state.

\begin{figure}[h!]
\begin{center}
\epsfig{file=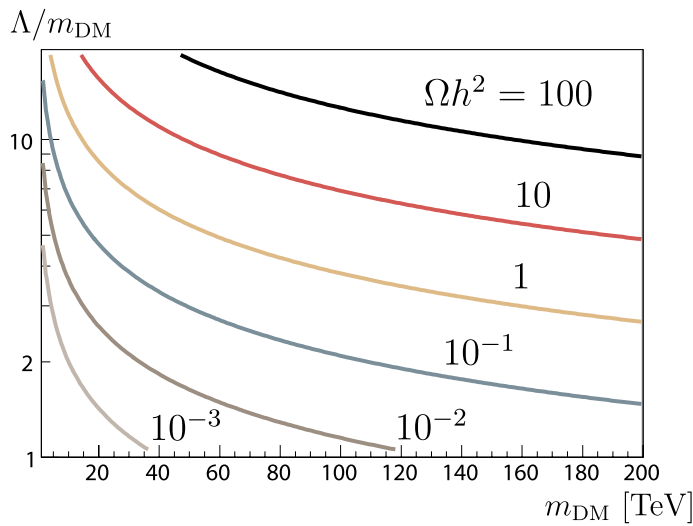 ,scale=.5,clip}
\end{center}
\caption{Contour plot of the DM abundance as a function of the DM mass and $\Lambda/m_{\rm DM}$.
 }
\label{fig:Abundance}
\end{figure}

In Fig.~\ref{fig:Abundance}, we show the contour plot of the DM abundance 
as a function of the DM mass and $\Lambda/m_{\rm DM}$.
Here, for simplicity, the result is given for the supersymmetric case,
following Eq.~(\ref{ann_cross_sec_NG}).
We set $2g_M=g_B=4\pi$ and $d=48$ in  Eq.~(\ref{ann_cross_sec_NG}) 
\footnote{$d=48$ comes from the degrees of freedom of the NG bosons and its scalar and fermionic partners.
In fact, the scalar and fermionic partners of the NG bosons
interact with the DM in a different manner from the NG bosons, which leads
to the slightly different cross section from Eq.~(\ref{ann_cross_sec_NG}).
However, its difference does not change the following result essentially.
} 
and neglect the meson mass, since the meson mass is
smaller than the DM mass; 
typically $m_{\rm ad}\simeq (g_M/g_B)^2 m_{\rm DM} $.

\section{Composite gauge mediation}

In our composite messenger gauge mediation model, the supersymmetry breaking
is mediated dominantly by the ${\rm SU}(5)_{\rm GUT}$-adjoint messenger meson
when $m=h\langle S\rangle<\Lambda$.
In this section, we calculate
the supersymmetry-breaking soft masses for the MSSM gauginos and scalars
based on the method in ref. \cite{GR}.

The method of ref. \cite{GR} is to consider the effective coupling of the goldstino superfield
to the MSSM fields. It is obtained by first considering the supersymmetric case and integrating
out the messenger sector, and then analytically continuating the messenger mass
to the goldstino superfield.
In the present model, the goldstino field is given by a linear combination of $\delta S$, $\delta X$ and $Z$,
where $Z$ is some singlet field in the SUSY breaking sector.

Let us first consider the gaugino masses. The gaugino masses are obtained by
considering the effective gauge kinetic term after the integration out of
the messenger sector:
\begin{equation}
{\cal L}\supset\frac{1}{8\pi}{\rm Im}\left[\int d^2\theta\,\tau(\Phi_0,Q)W^aW^a\right],
\end{equation}
where we denote the goldstino field by $\Phi_0$ and $Q$ is the renormalization scale. In the present model,
\begin{equation}
\Phi_0\propto\frac{g_X}{16\pi^2}\Lambda^2(v^5-1)\delta X-\frac{\mu^2}{2|f|}
\delta S + F_Z Z.
\end{equation}
Then the gaugino mass is given by
\begin{equation}
m_\lambda(Q)=-\frac{1}{2}\left.\frac{\partial{\rm log}\,\tau(\Phi_0,Q)}{\partial{\rm log}\,\Phi_0}\right|_{\Phi_0=\langle\varphi_0\rangle}
\frac{F_0}{\langle\varphi_0\rangle},
\end{equation}
where $\varphi_0$ and $F_0$ are the scalar- and $F$-component of the $\Phi_0$, respectively. In our case, however,
the same result can be obtained by calculating
\begin{equation}
m_\lambda(Q)=-\frac{1}{2}\left.\frac{\partial{\rm log}\,\tau(S,Q)}{\partial{\rm log}\,S}\right|_{S=\langle S\rangle}
\frac{F_S}{\langle S\rangle},
\end{equation}
because of the relations in Eqs.~(\ref{scalar_rel}) and (\ref{F_rel}).

The effective gauge coupling $\tau(S,Q)$ is obtained by considering the threshold correction of the adjoint messenger meson.
In the ${\rm SU}(5)_{\rm GUT}$-invariant case, the effective gauge coupling in the supersymmetric limit is given by
\begin{equation}
\tau(Q)=\tau(Q_{\rm UV})-\frac{b_{\rm H}}{2\pi i}{\rm log}\left(\frac{|m_{\rm ad}|}{Q_{\rm UV}}\right)
-\frac{b_{\rm L}}{2\pi i}{\rm log}\left(\frac{Q}{|m_{\rm ad}|}\right),
\end{equation}
where $b_{\rm H}$ and $b_{\rm L}$ are the one-loop coefficients of the beta function with and without
the adjoint meson, respectively, and we have $b_{\rm L}-b_{\rm H}=5$.
By holomorphy, we have
\begin{equation}
\tau(S,Q)=\frac{b_{\rm L}-b_{\rm H}}{2\pi i}{\rm log}(hS)+(\text{terms not involving $S$}).
\end{equation}
The gaugino mass at one-loop order is then obtained as
\begin{equation}
m_\lambda(Q)=5\frac{\alpha(Q)}{4\pi}\frac{F_S}{\langle S\rangle}.
\end{equation}
In the case that the ${\rm SU}(5)_{\rm GUT}$ is spontaneously broken down to the SM gauge group,
different mesons contribute threshold corrections to different gauge group factors. However,
even in this case the same equation holds for each factor gauge group
because of the linear dependence of the meson mass on $\langle S\rangle$.

Next, we consider the scalar masses. The scalar soft masses are obtained from the
wave-function renormalization of the MSSM matter superfield:
\begin{equation}
{\cal L}\supset\int d^4\theta\,Z_i(\Phi_0,\Phi_0^*,Q)\Phi_i^\dagger\Phi_i,
\end{equation}
where $\Phi_i$ is one of the MSSM matter superfield and $i$ is an index for each irreducible representation
of the SM gauge group. Then the soft masses are calculated as
\begin{equation}
\tilde{m}_i^2=-\left.\frac{\partial^2{\rm log}\,Z_i(Q)}{\partial{\rm log}\,\Phi_0\partial{\rm log}\,\Phi_0^*}\right|_{\Phi_0=\varphi_0}
\left|\frac{F_0}{v_0}\right|^2.
\end{equation}
Here, by the same reason as above, $\Phi_0$ in this  expression can be replaced by $S$.
As discussed in ref. \cite{GR}, the leading contribution can be calculated by the one-loop anomalous dimension.
The threshold correction of the messengers to the one-loop anomalous dimension is
\begin{equation}
Q\frac{d}{dQ}Z_i(Q)=\sum_{a=1}^3 \frac{C_a^i}{\pi}\alpha_a,
\end{equation}
where $\alpha_1,\alpha_2,\alpha_3$ denote the SM gauge couplings and $C_a^i$ is the quadratic Casimir of the $\Phi_i$
for the gauge group $a=1,2,3$. The result is
\begin{equation}
\tilde{m}_i^2=\sum_{a=1}^32C_a^i\left(\frac{\alpha_a}{4\pi}\right)^2\left|\frac{F_S}{\langle S\rangle}\right|^2.
\end{equation}
The higher-order corrections to the leading result for the soft masses are
suppressed by factors of $F/m^2$ or $m/\Lambda$.

\subsection*{MSSM Mass Spectrum}
\begin{figure}[h!]
\begin{center}
\epsfig{file=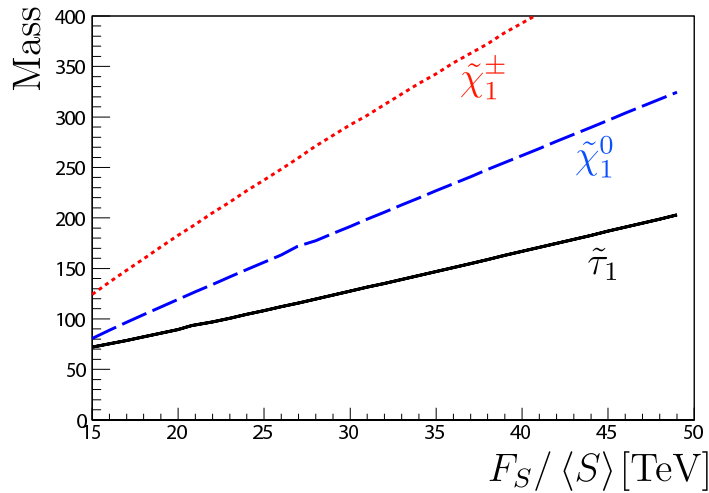 ,scale=.5,clip}
\end{center}
\caption{Masses of the lightest stau, neutralino, and chargino as functions of $F_S/\left<S\right>$.
The gluino mass is about five times larger than the lightest neutralino mass.}
\label{fig:GM}
\end{figure}

The low energy mass spectrum is almost the same as that of minimal GMSB, except for the unknown higher  $F/m^2$ and $m/\Lambda$ terms.
In Fig.~\ref{fig:GM}, we show some masses of MSSM particles.
Here we have used the program {\verb+SOFTSUSY+}~\cite{Allanach:2001kg},
setting the messenger mass scale to be $2F_S/\left<S\right>$ and $\tan\beta=10$.
The mass of messenger itself dose not affect the MSSM spectrum very much. 
We can see that $F_S/\left<S\right> \gsim 25$ TeV must be satisfied to evade the constraint from the collider experiments.

\section*{An Example}
Here, we pick out a model point which generates correct dark matter abundance and MSSM sparticle masses.
If we set $\Lambda=250$ TeV, $\mu=45$ TeV, $h=0.8$,  $f=0.25$ and $g_X = g_B = 2g_M = 4\pi$,
we have $m_{\rm DM}=101$ TeV and annihilation cross section 
$\sigma v_{\rm rel}\simeq 2\times 10^{-9}~\GEV^{-2}$, following Eq.~(\ref{ann_cross_sec_NG}).
The mass and annihilation cross section satisfy the constraint from the unitarity bound Eq.~(\ref{eq:UpDM}).
As for the MSSM mass spectrum, we have $F_S/\left<S\right> = 27$ TeV, which corresponds to
$m_{\tilde{\tau}} \simeq 100$ GeV.
This MSSM mass spectrum is consistent with the current experimental bounds.

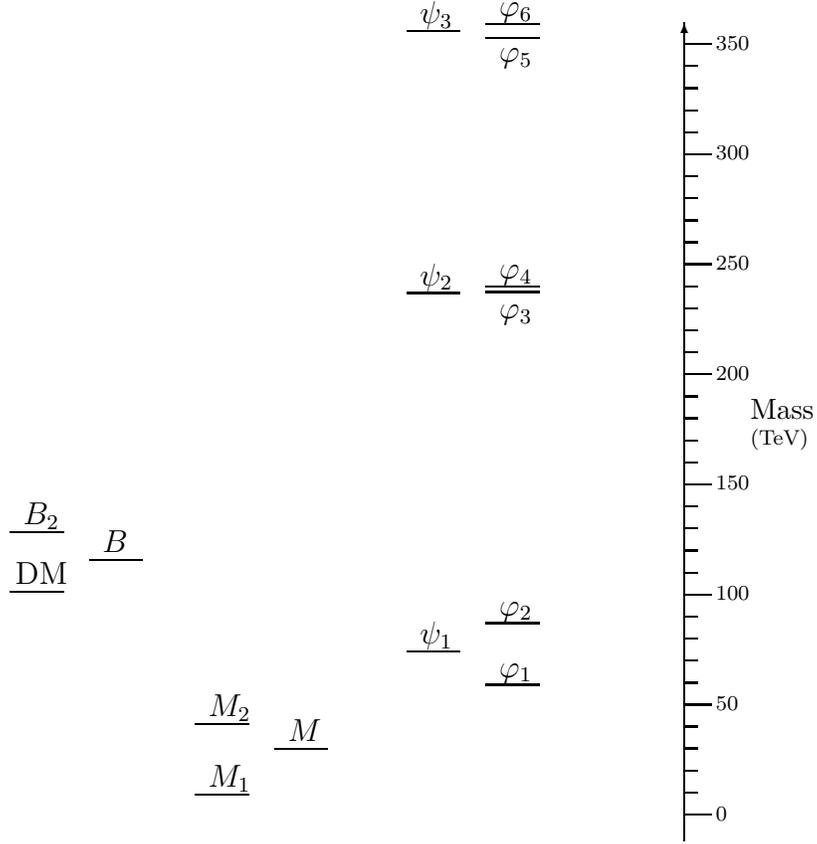
\begin{figure}[h!]
\begin{center}
\input{out}
\end{center}
\caption{Mass spectrum of the messenger sector at the sample point.
$\psi_i$ and $\varphi_i$ are the singlet fermions and scalars, respectively.}
\label{fig:MASS}
\end{figure}

The mass spectrum in the messenger sector for this parametrization is illustrated in {Fig.~\ref{fig:MASS}}.

\section{Discussion and Conclusions}
Let us discuss the decays of the messenger mesons $M^i_j$. 
When the ${\rm SU}(5)_{\rm GUT}$ group is broken down to the SM gauge group,
the mesons split into the octet (the adjoint of ${\rm SU}(3)$), the triplet (the adjoint of ${\rm SU}(2)$),
the hybrid (the $({\bf3},{\bf2})$ representation of ${\rm SU}(3){\times}{\rm SU}(2)$), and two singlets.
The color octet, weak triplet and singlet mesons can decay into a pair of gauge multiplets in the SSM 
and hence the decay is very fast. However, the hybrid mesons, $M_{{\bar d}{\bar \ell}}$ and $M_{\ell d}$,
decay only through non-renormalizable interactions. The dominant interaction is given by
\begin{equation}
W = \frac{1}{m_*}Q^d{\bar Q}_\ell H_u{\bar d},
\end{equation}
where $m_*$ denotes some high energy scale (e.g. the Planck scale).
In the effective theory of hadrons, the interaction and the mass term for the hybrid meson are written as
\begin{equation}
W=\frac{g_M}{16\pi^2}\frac{\Lambda}{m_*}M_{d\ell}H_u\bar{d}+m_{\rm hyb}M_{d\ell}M_{\bar{d}\bar{\ell}},
\end{equation}
and the decay rate is given by
\begin{equation}
\Gamma\simeq\frac{1}{16\pi}\left(\frac{g_M}{16\pi^2}\right)^2\frac{\Lambda^2 m_{\rm hyb}}{m_*^2}.
\end{equation}
For example, if $\Lambda=100$ TeV, $m_{\rm hyb}=50$ TeV, and $m_*=10^{16}$ GeV, we have
$\Gamma\simeq(10^{-2}~{\rm sec})^{-1}$,
and hence they become cosmologically safe.

In this paper, we have discussed the possibility of the dark matter of the composite messenger in GMSB models.
We show that there is a stable composite baryon which has proper annihilation 
cross section for the present dark matter abundance
and the successful gauge mediation can be achieved.

We have treated down-type and lepton-type messenger with no distinction.
However, the difference of the two types of the messengers can be important in some case \cite{Hamaguchi:2008rv}.
We should also mention the dark matter decay through higher dimensional operators.
For a dimension six operator,
the composite dark matter can have very long but finite lifetime.
The anomalies of recent cosmic ray experiments can be explained by the dark matter decay \cite{Hamaguchi:2008rv}.
Detailed analysis will be done in a forthcoming paper \cite{NS}.

\subsection*{Acknowledgement}
This work was supported by World Premier International Center Initiative (WPI Program),
MEXT, Japan.
The work of K.H. was supported by JSPS Grant-in-Aid for Young Scientists (B) (21740164).
The work of E.N. and S.S. is supported in part by JSPS Research Fellowships for Young Scientists.

\end{document}

%% file: out.tex
\scalebox{1}{
\begin{picture}(220,300)(0,0)
\put(5,107.023){\line(1,0){20}}
\put(10,110.023){$ { B_2}$}
\put(5,84.4143){\line(1,0){20}}
\put(7,87.4143){DM}
\put(35,96.384){\line(1,0){20}}
\put(40,99.384){$ {B}$}

\put(75,34.243){\line(1,0){20}}
\put(80,37.243){$ M_2$}
\put(75,7.64692){\line(1,0){20}}
\put(80,10.6469){$ M_1$}
\put(105,24.8098){\line(1,0){20}}
\put(110,27.8098){$ M$}

\put(155,61.6667){\line(1,0){20}}
\put(160,64.6667){$ \psi_{1}$}
\put(155,197.5){\line(1,0){20}}
\put(160,200.5){$ \psi_{2}$}
\put(155,296.667){\line(1,0){20}}
\put(160,299.667){$ \psi_{3}$}

\put(185,49.1667){\line(1,0){20}}
\put(190,52.1667){$ \varphi_{1}$}
\put(185,72.5){\line(1,0){20}}
\put(190,75.5){$ \varphi_{2}$}
\put(185,197.833){\line(1,0){20}}
\put(190,187.833){$ \varphi_{3}$}
\put(185,200){\line(1,0){20}}
\put(190,203){$ \varphi_{4}$}
\put(185,294.167){\line(1,0){20}}
\put(190,285.167){$ \varphi_{5}$}
\put(185,299.167){\line(1,0){20}}
\put(190,302.167){$ \varphi_{6}$}
\put(285,150){\small Mass }
\put(285,140){$\scriptstyle {\rm (TeV)}$ }
\put(260,-10){\vector(0,1){310}}
\put(260,0){\line(1,0){10}}
\put(272,-2){$\scriptstyle 0$}
\put(260,41.6667){\line(1,0){10}}
\put(272,39.6667){$\scriptstyle 50$}
\put(260,83.3333){\line(1,0){10}}
\put(272,81.3333){$\scriptstyle 100$}
\put(260,125){\line(1,0){10}}
\put(272,123){$\scriptstyle 150$}
\put(260,166.667){\line(1,0){10}}
\put(272,164.667){$\scriptstyle 200$}
\put(260,208.333){\line(1,0){10}}
\put(272,206.333){$\scriptstyle 250$}
\put(260,250){\line(1,0){10}}
\put(272,248){$\scriptstyle 300$}
\put(260,291.667){\line(1,0){10}}
\put(272,289.667){$\scriptstyle 350$}
\put(260,0){\line(1,0){5}}
\put(260,8.33333){\line(1,0){5}}
\put(260,16.6667){\line(1,0){5}}
\put(260,25){\line(1,0){5}}
\put(260,33.3333){\line(1,0){5}}
\put(260,41.6667){\line(1,0){5}}
\put(260,50){\line(1,0){5}}
\put(260,58.3333){\line(1,0){5}}
\put(260,66.6667){\line(1,0){5}}
\put(260,75){\line(1,0){5}}
\put(260,83.3333){\line(1,0){5}}
\put(260,91.6667){\line(1,0){5}}
\put(260,100){\line(1,0){5}}
\put(260,108.333){\line(1,0){5}}
\put(260,116.667){\line(1,0){5}}
\put(260,125){\line(1,0){5}}
\put(260,133.333){\line(1,0){5}}
\put(260,141.667){\line(1,0){5}}
\put(260,150){\line(1,0){5}}
\put(260,158.333){\line(1,0){5}}
\put(260,166.667){\line(1,0){5}}
\put(260,175){\line(1,0){5}}
\put(260,183.333){\line(1,0){5}}
\put(260,191.667){\line(1,0){5}}
\put(260,200){\line(1,0){5}}
\put(260,208.333){\line(1,0){5}}
\put(260,216.667){\line(1,0){5}}
\put(260,225){\line(1,0){5}}
\put(260,233.333){\line(1,0){5}}
\put(260,241.667){\line(1,0){5}}
\put(260,250){\line(1,0){5}}
\put(260,258.333){\line(1,0){5}}
\put(260,266.667){\line(1,0){5}}
\put(260,275){\line(1,0){5}}
\put(260,283.333){\line(1,0){5}}
\put(260,291.667){\line(1,0){5}}
\end{picture} }

%% file: CMDM.bbl
\begin{thebibliography}{99}

\bibitem{Griest:1989wd}
  K.~Griest and M.~Kamionkowski,
  Phys.\ Rev.\ Lett.\  {\bf 64}, 615 (1990).

\bibitem{Dimopoulos:1996gy}
  S.~Dimopoulos, G.~F.~Giudice and A.~Pomarol,
  Phys.\ Lett.\  B {\bf 389}, 37 (1996)
  [arXiv:hep-ph/9607225].

\bibitem{Hamaguchi:2007rb}
  K.~Hamaguchi, S.~Shirai and T.~T.~Yanagida,
  Phys.\ Lett.\  B {\bf 654}, 110 (2007)
  [arXiv:0707.2463 [hep-ph]].

\bibitem{Hamaguchi:2008rv}
  K.~Hamaguchi, E.~Nakamura, S.~Shirai and T.~T.~Yanagida,
  Phys.\ Lett.\  B {\bf 674}, 299 (2009)
  [arXiv:0811.0737 [hep-ph]].

\bibitem{G1}
  H.~Pagels and J.~R.~Primack,
  Phys.\ Rev.\ Lett.\  {\bf 48} (1982) 223;
\\
  M.~Viel, J.~Lesgourgues, M.~G.~Haehnelt, S.~Matarrese and A.~Riotto,
  Phys.\ Rev.\  D {\bf 71} (2005) 063534
  [arXiv:astro-ph/0501562].

\bibitem{G2}
  K.~Kohri, T.~Moroi and A.~Yotsuyanagi,
  Phys.\ Rev.\  D {\bf 73} (2006) 123511
  [arXiv:hep-ph/0507245].

\bibitem{NS} E. Nakamura and S. Shirai, in preparation.

\bibitem{Seiberg:1994bz}
  N.~Seiberg,
  Phys.\ Rev.\  D {\bf 49}, 6857 (1994)
  [arXiv:hep-th/9402044].

\bibitem{Luty}
  M.~A.~Luty,
  Phys.\ Rev.\  D {\bf 57} (1998) 1531
  [arXiv:hep-ph/9706235];
\\
  A.~G.~Cohen, D.~B.~Kaplan and A.~E.~Nelson,
  Phys.\ Lett.\  B {\bf 412} (1997) 301
  [arXiv:hep-ph/9706275].

\bibitem{Fujii:2003iw}
  M.~Fujii, M.~Ibe and T.~Yanagida,
  Phys.\ Rev.\  D {\bf 69}, 015006 (2004)
  [arXiv:hep-ph/0309064].

\bibitem{Witten:1982df}
  E.~Witten,
  Nucl.\ Phys.\  B {\bf 202}, 253 (1982).

\bibitem{Komatsu:2008hk}
  E.~Komatsu {\it et al.}  [WMAP Collaboration],
  Astrophys.\ J.\ Suppl.\  {\bf 180}, 330 (2009)
  [arXiv:0803.0547 [astro-ph]].

\bibitem{GR}
  G.~F.~Giudice and R.~Rattazzi,
  Nucl.\ Phys.\  B {\bf 511}, 25 (1998)
  [arXiv:hep-ph/9706540].

\bibitem{Allanach:2001kg}
  B.~C.~Allanach,
  Comput.\ Phys.\ Commun.\  {\bf 143}, 305 (2002)
  [arXiv:hep-ph/0104145].






\end{thebibliography}
